\documentclass[aps,pra,article,twocolumn]{revtex4}
\usepackage{dcolumn}
\usepackage{bbm}
\usepackage{epsfig,psfrag}
\usepackage{t1enc}
\usepackage[latin1]{inputenc}
\usepackage{graphicx}
\usepackage[dvips]{rotating}
\usepackage{flafter}
\usepackage{amssymb}
\usepackage{amsmath}
\usepackage{pifont}
\usepackage{float}
\usepackage{latexsym}

\begin{document}


\title{Simulation of colloidal chain movements under a magnetic field}

\author{Mareike Gr\"unzel}

\affiliation{Institut f\"ur Theoretische Physik, Universit\"at zu
K\"oln, Z\"ulpicher Stra\ss e 77, D-50937 K\"oln, Germany}

\date{\today}

\begin{abstract}
Short colloidal chains are simulated by the slithering-snake-algorithm on a simple cubic lattice. The dipole character of the colloidal particles leads to a long range dipole-dipole interaction. The solvent is simulated by the nearest neighbor Ising model. The aligning of the dipoles under a magnetic field gives rise to the chains to align on their part with the field direction. 
\end{abstract}

\maketitle


\section{Introduction}
The initial idea of this paper is based on the article of Goubault et al. \cite{paper}, whose research partly dealt with the aligning of colloidal chains and their curvature behavior being influenced by an external magnetic field.
\\
The basis of the program is another program that simulates short polymer chains with a hydrophilic head, a hydrophobic tail and a neutral center surrounded by oil and water molecules, which are simulated with the nearest neighbor spin Ising model \cite{paperstauffer}. The movements of the chains are simulated by the slithering-snake algorithm. This program was created on the basis of Larson's PhD thesis \cite{larson}. 

\section{Aligning of the chains}\label{aligning}
The Ising model and the slithering-snake algorithm are taken over to the new program. The first aim of the work is to add the long-range dipole-dipole interaction according to the colloidal character of the chains to the program. The magnetic dipole interaction energy for two interacting magnetic momenta $\vec\mu_{i}$ and $\vec\mu_{j}$ (e.g. \cite{Jackson}) is transformed to scalar spin variables $s_i=\pm 1$ so that this leads to the Hamiltonian \cite{diplom}
\begin{displaymath}\label{HamiltonDipolneu}
{\cal H}=-\frac{1}{2}\sum_{i,j}\,J_{i,j}\,s_i\,s_j\quad\text{ with}
\end{displaymath}
\begin{displaymath}
J_{i,j}=
\begin{cases}
J -(g/r_{ij})^{3}\cdot\left(1-3\cdot(r_{ij}^{z}/r_{ij})^{2}\right) & i,j\text{ neighbors,}
\\
-(g/r_{ij})^{3}\cdot\left(1-3\cdot(r_{ij}^{z}/r_{ij})^{2}\right) &\text{else},
\end{cases}
\end{displaymath}
where $g=\mu^2/a^3$ is the dipole strength, $r_{ij}^{z}$ is the 
component of the vector $\vec r_{ij}=(r_{ij}^x,r_{ij}^y,r_{ij}^z)$ between 
two regarded dipoles and $r_{ij}$ is the distance between them. The space 
coordinate $i$ varies from 1 to $L^3$ for $L \times L \times L$ lattices. The 
calculation of the distance is being done by the calculation of the 
coordinates of the two regarded dipoles with the help of the 
modulo-function. The x-coordinates, for example, are the modulo-function 
of the positions and the system width $L$. The difference between the x-, y- 
and z-coordinates gives the distance vector $\vec r_{ij}$ of the dipoles. 

Initially, the chains are aligned in the x-direction 
in every simulation. The magnetic field is oriented in z-direction.
\begin{figure}[ht]
\centerline{\psfig{file={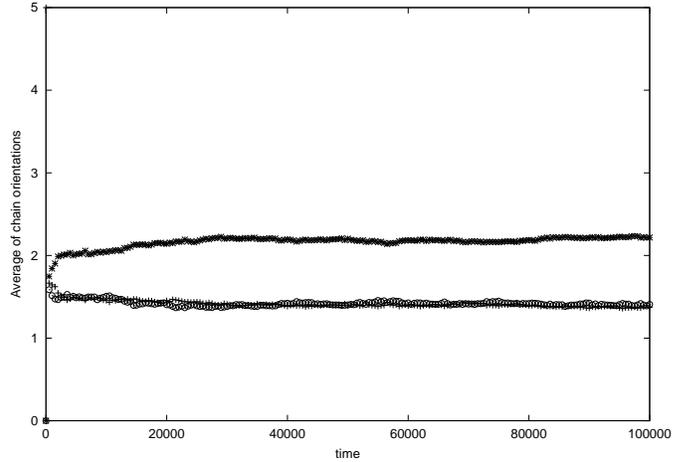}, scale=0.35, angle=270}}
\caption{Average of chain aligning of 5 chains of length 6 (concentration of 0.03) in a $10\times 10\times 10$ lattice at the temperature of $3.0\,J/k_B$ (100 runs). (+): x-aligning; ({\Large$\circ$}): y-aligning; (\ding{83}): z-aligning.}
\label{L10_len6_np5_max100_Zeit100000_23}
\end{figure}
The computation of the energy is summed up for one chain interacting with the entire system due to the long range of the interaction. This is repeated for every chain. The spins of the chain molecules range from value $s_i=-1$ for the tail, over to $s_i=0$ for the neutral center, and $s_i=1$ for the second molecule over to $s_i=2$ for the head.
\\
In this case, the magnetic field is simulated indirectly by aligning the 
dipoles in the direction of the field. The chains align in the field 
direction until equilibrium is achieved after approximatly 30000 time 
steps to be seen in Fig. \ref{L10_len6_np5_max100_Zeit100000_23}, where 
the average orientations of 5 chains of length 6 in a $10\times 10\times 
10$ 
lattice is shown. 

The first method to calculate the aligning is the 
determination of the distances of the head and the tail of the chains. 
This method favors finite-size effects in smaller systems, because the 
chains can spread over the lattice. - That is why only the second method 
has been used. It calculates the difference between two chain links 
situated next to each other. Then it is being decided in which direction 
they are standing to each other. If the difference is $\pm 1$, the 
x-aligning is raised  by 1; if the difference is $\pm L$, 
the y-aligning is raised and if it is $\pm L^2$ it concerns to the 
z-aligning. The average is calculated for every chain and every run 
of the program at every timestep to get the course of chain aligning 
in dependence on the time. Taking the example of a chain with 
the length of 6, it can be arranged at a maximum of 5 in one direction. 
\\
Simulations with a variation of the concentration of the chains result in a faster rise of the average alignments towards the field direction for small concentrations. 
\begin{figure}[ht]
\centerline{\psfig{file={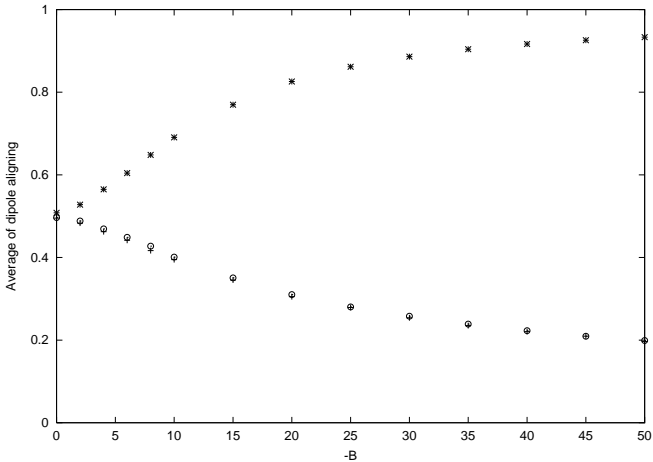}, scale=0.35, angle=270}}
\caption{Average of dipole aligning of 5 chains of length 6 (concentration 0.03) in a $10\times 10\times 10$ lattice at the temperature of $3.0\,J/k_B$ (100 runs for each magnetic field strength). (+): x-component; ({\Large$\circ$}): y-component; (\ding{83}): z-component.}
\label{dipolminbfeld}
\end{figure}

\section{Modified model}
The program has a modified version, in which the dipoles are three-dimensional and rotate randomly. The rotation takes place with the method of Barker \& Watts in their simulation of water (1969) \cite{CompSimofLiquids}. Molecules rotate with a random angle between zero and a specified maximum angle (in this case $\frac{\pi}{2}$) to all sides about one space-fixed axes chosen at random. The Hamiltonian is now changed as follows: 
\begin{displaymath}\label{HamiltonDipolBfeld}
{\cal H}=-\frac{1}{2}\sum_{i,j} J_{i,j}-\sum_{i} \vec B \cdot \vec s_i \quad
\text{ with}
\end{displaymath}
\begin{displaymath}
J_{i,j}=
\begin{cases}
J -(g/r_{ij}^{3})\left(\vec s_i\,\vec s_j - 3 \left(\vec s_i\cdot\vec r_{ij}\right)\left(\vec s_j\cdot\vec r_{ij}\right)/r_{ij}^{2}\right)
\\
-(g/r_{ij}^{3})\left(\vec s_i\,\vec s_j - 3 \left(\vec s_i\cdot\vec r_{ij}\right)\left(\vec s_j\cdot\vec r_{ij}\right)/r_{ij}^{2}\right)
\end{cases}
\end{displaymath}
After the calculation of the energy to align the chains there is another 
calculation for every dipole of the chains relative of the rotation of the 
dipoles. The dipole strength has the value 2 for the head of every chain 
and 1 for the rest in all modified simulations. The magnetic field now 
influences directly the rotation of the dipoles and therefore indirectly 
the aligning of the chains in the direction of the field. The magnetic 
field is negative in all simulations due to the fact, that a negative 
field affects the dipoles to rotate in the positive z-direction, because 
they are effective against the magnetic field. 
Simulations with a positive magnetic field result in a suppression of the aligning in the direction of the field.
Instead, simulations with a negative magnetic field show the expected 
behavior of the dipoles under a magnetic field with various field 
strengths (Fig. \ref{dipolminbfeld}). In Fig. \ref{dipolminbfeld} the 
average value of each component is computed for 70 timesteps from 30 to 
100 for every magnetic field strength. This shows the equal arrangement of 
the dipoles without a magnetic field in a nice way, because every dipole 
component has the equal value $0.5$. Simulations 
result in a constant value after 30 timesteps creating a sufficient average value. The errors of these average values are too small to be visible.
Simulations with the dipole aligning not random but in x-direction for the initial condition result in the same behavior.
\begin{figure}[ht]
\centerline{\psfig{file={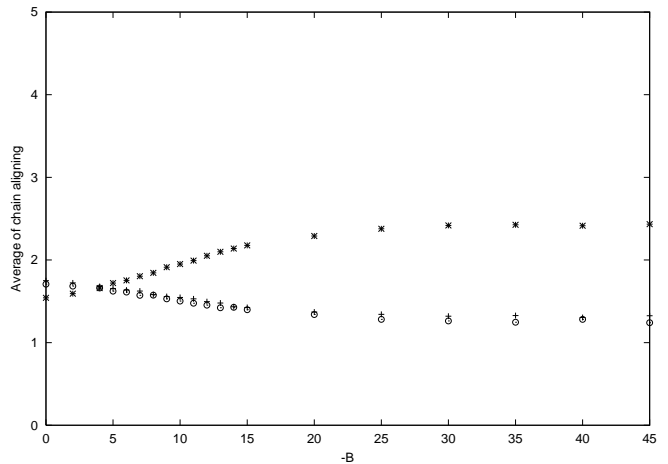}, scale=0.35, angle=270}}
\caption{Average of the chain aligning of 5 chains of length 6 (concentration 0.03) in a $10\times 10\times 10$ lattice at the temperature of $3.0\,J/k_B$ (100 runs for each magnetic field strength). (+): x-aligning; ({\Large$\circ$}): y-aligning; (\ding{83}): z-aligning.}
\label{kettenminbfeld}
\end{figure}

Fig. \ref{kettenminbfeld} shows the chain aligning of 5 chains depending on 
the magnetic field strength. The average value is computed for 850 
timesteps from 150 to 1000. As Fig. \ref{L10_len6_np5_max100_Zeit100000_23} shows, this duration is not enough to complete the aligning, but its tendency is shown in Fig. \ref{kettenminbfeld}. For the interval from $B=0$ to $B=3$ the aligning in the direction of the field does not take place due to the high temperature of $T=3.0\,J/k_B$. For high magnetic field strengths there is a saturation of the magnetic field, so that a rise of the aligning is no longer possible.
Instead, the middle interval of Fig. \ref{kettenminbfeld} shows a linear process of aligning in direction of the field. 
The gradient of the aligning in x- and y-direction is nearly the same. The line pointing into the x-direction, however, is situated a bit higher than the one pointing into the y-direction, because the chains have not arranged yet due to the short time; in the beginning they were arranged into the x-direction. Despite the small time for the average value, this is a clear tendency. 
\\
Simulations with a discrete (random angle of $\pi/2$ or $-\pi/2$) aligning 
result in the same behavior but not as exact as it is in the continuous 
aligning. The discrete model shows various weak points in contrast to the 
continuous case. Among other things, the dipole aligning takes place much 
faster to the value 1. In relation to the chain aligning, there is no 
visible linear dependence on the magnetic field.

\section{Conclusion}
The result is a good working program offering lots of possibilities to 
work with. Among other things, the dependence on the temperature or the 
chain length and concentration can be examined. The results from Goubault 
et. al. \cite{paper} can likewise be researched with little effort, above 
all the curvature of the chains. The easiest way to examine this is taking 
the first method, which has been mentioned in section \ref{aligning}. For 
this, bigger systems are necessary. The linear dependence of 
the aligning on the magnetic field should be verified for bigger systems.

\end{document}